\documentstyle[prl,aps,epsf]{revtex}
\begin{document}
\title{Theory of the angular magnetoresistance in CPP spin valves}
\author{Daniel Huertas-Hernando, Gerrit E. W. Bauer and Yu. V. Nazarov}
\address{Department of Applied Physics and Delft Institute of Microelectronics and
Submicrontechnology, \\
Delft University of Technology, Lorentzweg 1, 2628 CJ Delft, The Netherlands}
\date{\today}
\maketitle

\begin{abstract}
The resistance of CPP spin valve is a continuous function of the angle $%
\theta $ between the magnetizations of both ferromagnets. We use the cicuit
theory for non-collinear magnetoelectronics to compute the angular
magnetoresistance of CPP spin valves taking the spin accumulation in the
ferromagnetic layers into account.
\end{abstract}

\bigskip 

The magnetoresistance of multilayers in the so-called current perpendicular
to the plane (CPP) configuration first studied by the Michigan State
University \ and Philips groups \cite{RefPratt}, is still an active field of
research. Most studies have been focused on collinear, parallel and
antiparallel, magnetic configurations. There are several papers that have
investigated also non-collinear transport \cite{RefSlonczewski,RefVedyayev}.
The circuit theory of mesoscopic transport \cite{RefYuli} has been applied
to describe transport in non-collinear magnetic structures \cite{RefCircuit}%
. Interesting phenomena like spin precession effects on the induced spin
accumulation have been found by solving the diffusion equation for CPP spin
valves \cite{RefDani}. An important concept in the theory of non-collinear
spin transport is the so-called {\em mixing conductance }$G^{\uparrow
\downarrow }$ \cite{RefCircuit} which is closely related to the
current-induced spin-torque \cite{RefKatine}. In this paper we extend our
previous theoretical work on the total conductance of CPP ferromagnet-normal
metal-ferromagnet (F%
%TCIMACRO{\TEXTsymbol{\vert}}%
%BeginExpansion
\mbox{$\vert$}%
%EndExpansion
N%
%TCIMACRO{\TEXTsymbol{\vert}}%
%BeginExpansion
\mbox{$\vert$}%
%EndExpansion
F) spin valves, taking into account the spin-accumulation in the
ferromagnets \cite{Fert:prb96}.

In our model two ferromagnetic layers (F) are connected to a normal layer
through metallic contact regions or interfaces (C). The system may split up
into reservoirs, resistors and nodes (Fig.1) \cite{Schep:prb97}. Two
fictitious nodes connect each F-layer with a contact region C, where F is
represented by the spin-up and spin-down conductances and C is described by
four spin-dependent conductances. We assume that no spin-flip processes
occur neither at the interfaces nor in the bulk layers{\bf . }The resistance
in the normal metal node is also disregarded. Associated with each node is a
2$\times $2 distribution matrix in spin space which represents charge and
spin-accumulation. The spin accumulation in the ferromagnetic (F-C) node is
taken to be parallel to the magnetization ${\bf \vec{m}}_{1/2}$ of the
ferromagnet{\bf . }The spin and charge currents are then completely
determined by the relative orientation of the magnetization directions in
the ferromagnets (${\bf \vec{m}}_{1}\cdot {\bf \vec{m}}_{2}=\cos \theta $),
the spin-conductances associated with the ferromagnetic layers ($%
G_{F}^{\uparrow }$, $G_{F}^{\downarrow }$) and the contact conductances ($%
G_{c}^{\uparrow }$, $G_{c}^{\downarrow }$, $G^{\uparrow \downarrow }$).
According to first-principles calculations\cite{RefXia} $%
%TCIMACRO{\func{Re}}%
%BeginExpansion
\mathop{\rm Re}%
%EndExpansion
G^{\uparrow \downarrow }\gg 
%TCIMACRO{\func{Im}}%
%BeginExpansion
\mathop{\rm Im}%
%EndExpansion
G^{\uparrow \downarrow }$ so in the following we disregard the imaginary
part. For the symmetric case, in which the conductances are equal for both
interfaces and also for both F-layers, the total conductance reads 
\begin{equation}
G^{T}(\theta )=\frac{1}{2}\left\{ \frac{\left[ G_{F}G_{c}^{\uparrow
}G_{c}^{\downarrow }+G_{c}G_{F}^{\uparrow }G_{F}^{\downarrow }\right] 2%
%TCIMACRO{\func{Re}}%
%BeginExpansion
\mathop{\rm Re}%
%EndExpansion
G^{\uparrow \downarrow }\left( 1+\cos \theta \right) +\left[
4G_{c}^{\uparrow }G_{c}^{\downarrow }G_{F}^{\uparrow }G_{F}^{\downarrow }%
\right] \left( 1-\cos \theta \right) }{\left[ G_{c}^{\uparrow
}G_{c}^{\downarrow }+G_{F}^{\uparrow }G_{F}^{\downarrow }+G_{F}^{\uparrow
}G_{c}^{\downarrow }+G_{F}^{\downarrow }G_{c}^{\uparrow }\right] 2%
%TCIMACRO{\func{Re}}%
%BeginExpansion
\mathop{\rm Re}%
%EndExpansion
G^{\uparrow \downarrow }\left( 1+\cos \theta \right) +\left[
G_{F}G_{c}^{\uparrow }G_{c}^{\downarrow }+G_{c}G_{F}^{\uparrow
}G_{F}^{\downarrow }\right] \left( 1-\cos \theta \right) }\right\}
\label{GT}
\end{equation}

Eq. (\ref{GT}) reduces to well-known collinear expressions for $\theta =0$
and $\theta =\pi $. When $G_{F}\rightarrow \infty $, Eq.(24) in Ref. %
\onlinecite{RefDani} $G^{T}(\theta )$ is recovered. In the absence of a
spin-dependent interface resistance, {\em i.e. }$P_{c}=0$ $\left(
G_{c}^{\uparrow }=G_{c}^{\downarrow }\right) $, $G^{T}(0)-G^{T}(\pi )$ is
governed exclusively by the spin accumulation of the bulk F-layers.

Using Eq. (\ref{GT}), we obtain the following expression for the normalized
angular-MR (NAMR)%
%TCIMACRO{
%\TeXButton{TeX field}{\begin{mathletters}%
%}}%
%BeginExpansion
\begin{mathletters}%
%
%EndExpansion
\begin{equation}
NAMR(\theta )=\left( \frac{G^{T}(\theta )-G^{T}(\theta =0)}{G^{T}(\theta
=\pi )-G^{T}(\theta =0)}\right) \left( \frac{G^{T}(\theta =\pi )}{%
G^{T}(\theta )}\right) =\frac{1-\cos ^{2}\theta /2}{\chi \cos ^{2}\theta /2+1%
}  \label{AMR}
\end{equation}
with 
\begin{equation}
\chi =%
%TCIMACRO{\func{Re}}%
%BeginExpansion
\mathop{\rm Re}%
%EndExpansion
\eta \left[ \frac{1}{1-p_{c}^{2}}+\frac{\left( G_{c}/G_{F}\right) }{%
1-p_{F}^{2}}\right] -1.  \label{Ki}
\end{equation}
%TCIMACRO{
%\TeXButton{TeX field}{\end{mathletters}%
%} }%
%BeginExpansion
\end{mathletters}%
%
%EndExpansion
From Eq. (\ref{AMR}), we see that the NAMR depends on the angle $\theta $
between the magnetization of both ferromagnets and on the parameter $\chi $,
which depends on the contact and F-layer conductance parameters. Similar
expression for the NMAR have been obtained using the Kubo formalism \cite
{RefVedyayev}. They use a different model that forbids the direct comparison
of the results. In their case, no resistance was associated with the
interfaces between layers. Rather, the effect arises from the finite
resistance of the normal layer and spin-dependent resistance of the
ferromagnets.{\bf \ }

The parameters of the 2-channel series resistor model for collinear
configurations have been accurately determined from CPP magnetoresistance
experiments. For future comparison of our expression with NAMR experimental
data, the quantities $p_{c},$ $p_{F}$ and $G_{c}/G_{F}$ can be expressed in
terms of these parameters $\beta ,$ $\gamma ,$ $\rho _{F},$ $AR^{\ast }$ as 
%TCIMACRO{
%\TeXButton{TeX field}{\begin{mathletters}%
%}}%
%BeginExpansion
\begin{mathletters}%
%
%EndExpansion
\begin{equation}
p_{c}=P_{c}/G_{c}=\gamma ;\;p_{F}=P_{F}/G_{F}=\beta  \label{2CSR_1}
\end{equation}
\begin{equation}
G_{c}/G_{F}=\frac{\rho _{F}}{AR^{\ast }}\frac{t_{[F]}}{1-\gamma ^{2}}
\label{2CSR_3}
\end{equation}
such that 
\begin{equation}
\chi =%
%TCIMACRO{\func{Re}}%
%BeginExpansion
\mathop{\rm Re}%
%EndExpansion
\eta \left[ \frac{1}{1-\gamma ^{2}}+\frac{\rho _{F}}{AR^{\ast }}\frac{t_{[F]}%
}{\left( 1-\beta ^{2}\right) \left( 1-\gamma ^{2}\right) }\right] -1,
\label{Ki_t[F]}
\end{equation}
%TCIMACRO{
%\TeXButton{TeX field}{\end{mathletters}%
%} }%
%BeginExpansion
\end{mathletters}%
%
%EndExpansion
which shows the explicit dependence of the NAMR on the thickness of the
F-layer $t[F]$. From Eq. (\ref{Ki_t[F]}) we obtain the following expression
for $%
%TCIMACRO{\func{Re}}%
%BeginExpansion
\mathop{\rm Re}%
%EndExpansion
\eta $%
\begin{equation}
%TCIMACRO{\func{Re}}%
%BeginExpansion
\mathop{\rm Re}%
%EndExpansion
\eta =\left( 1+\chi \right) \left( 1-\gamma ^{2}\right) \left[ 1+\frac{\rho
_{F}}{AR^{\ast }}\frac{t_{[F]}}{1-\beta ^{2}}\right] ^{-1}.  \label{Re_eta}
\end{equation}
So fitting NAMR data using Eq. (\ref{AMR}) and Eq. (\ref{Ki_t[F]}), would
provide a way to obtain experimentally the mixing conductance $G^{\uparrow
\downarrow }$. The dependence on $t_{[F]}$ reflects the spin-accumulation in
the bulk ferromagnet. In the presence of spin-flip scattering, $t_{[F]}$
should be taken as the minimum of the geometrical layer thickness and the
spin-flip diffusion length. This extra parameter can be eliminated in favor
of measurable quantities by expressing $A\Delta R^{T}=R^{T}(\theta =\pi
)-R^{T}(\theta =0)$ in terms of $\beta ,$ $\gamma ,$ $\rho _{F},$ $AR^{\ast
} $: 
\begin{equation}
A\Delta R^{T}=\frac{1}{G^{T}(\theta =\pi )}-\frac{1}{G^{T}(\theta =0)}=2%
\text{ }AR^{\ast }\text{ }\gamma ^{2}\frac{\left[ 1+\frac{\rho _{F}}{%
AR^{\ast }}\frac{t_{[F]}}{1-\beta ^{2}}\left( \frac{\beta }{\gamma }\right) %
\right] ^{2}}{\left[ 1+\frac{\rho _{F}}{AR^{\ast }}\frac{t_{[F]}}{1-\beta
^{2}}\right] }  \label{deltaR}
\end{equation}
If we assume that $\left( \beta /\gamma \right) =1$ (For Py/Cu/Py: $\beta _{%
\text{Py}}=0.76,$ $\gamma _{\text{Py/Cu}}=0.7,$ $\left( \beta /\gamma
\right) =1.08$ and for CoFe/Cu/CoFe $\beta _{\text{CoFe}}=0.65,$ $\gamma _{%
\text{CoFe/Cu}}=0.75,$ $\left( \beta /\gamma \right) =0.86$) we can write%
%TCIMACRO{
%\TeXButton{TeX field}{\begin{mathletters}%
%} }%
%BeginExpansion
\begin{mathletters}%
%
%EndExpansion
\begin{equation}
\left[ 1+\frac{\rho _{F}}{AR^{\ast }}\frac{t_{[F]}}{1-\beta ^{2}}\right]
^{-1}=\frac{2\text{ }AR^{\ast }\text{ }\gamma ^{2}}{\text{ }A\Delta R^{T}}
\label{dR_1}
\end{equation}
and 
\begin{equation}
%TCIMACRO{\func{Re}}%
%BeginExpansion
\mathop{\rm Re}%
%EndExpansion
\eta =\left( 1+\chi \right) \left( 1-\gamma ^{2}\right) \frac{2\text{ }%
AR^{\ast }\text{ }\gamma ^{2}}{\text{ }\Delta R^{T}}  \label{Ren_dR}
\end{equation}
%TCIMACRO{
%\TeXButton{TeX field}{\end{mathletters}%
%}}%
%BeginExpansion
\end{mathletters}%
%
%EndExpansion
which gives an expression to determine Re$G^{\uparrow \downarrow }$ without
reference to the thickness parameter $t_{[F]}$.

In conclusion, the normalized angular magnetoresistance (NAMR) for CPP spin
valves depends on the spin accumulation in the F-layers. This effect should
be included in order to obtain reliable values of the mixing conductance by
fitting experimental NAMR\ data \cite{RefExpBill}.

We thank W. P. Pratt for useful discussions. This work is part of the
research program for the ``Stichting voor Fundamenteel Onderzoek der
Materie'' (FOM).

\begin{center}
{\bf Figure Caption}
\end{center}

{\bf Fig. 1: }Circuit{\bf \ }model of a CPP spin valve, in term of
reservoirs, resistors and nodes. Ferromagnetic layers (F) and metallic
contact regions (C) are represented by resistors, characterized by spin
conductance parameters. Two nodes connect each F-layer with a contact region
C. The third node represents the normal layer. Associated with each node is
a 2$\times $2 distribution matrix in spin space. The direction of the spin
accumulation in each magnetic node (F-C node) is denoted by a unit vector $%
{\bf \vec{m}}$. The external reservoirs (R) are supposed to be in local
equilibrium.

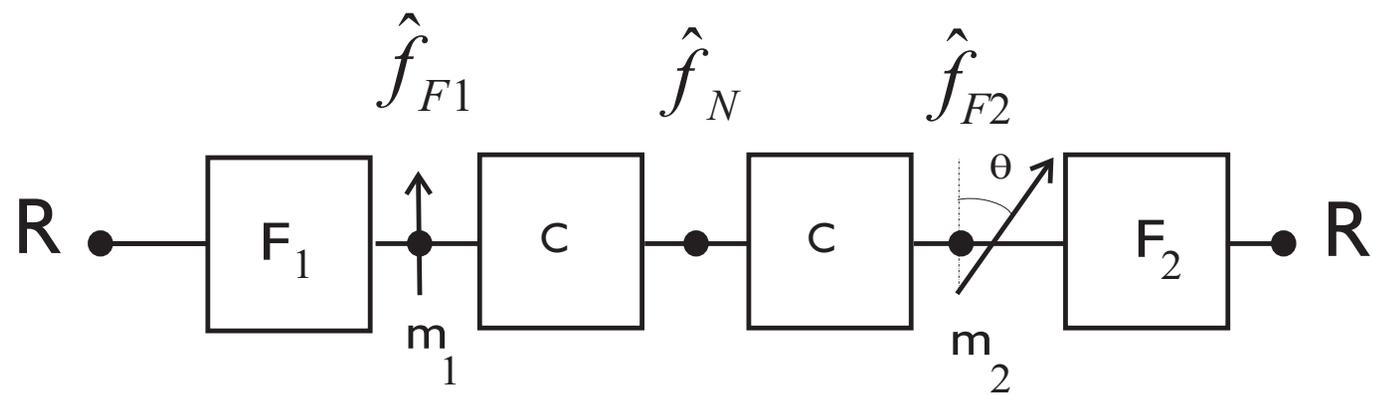
\end{document}